\documentstyle[12pt,epsfig]{article}
\newcommand{\z}{&&\hspace*{-1cm}}

\newcommand{\vph}{\varphi}
\newcommand{\bea}{\begin{eqnarray}}
\newcommand{\eea}{\end{eqnarray}}

\begin{document}

\begin{center}
{\Large \bf 
The structure functions $F_2^c$ and $F_L^c$\\
in the framework of the $k_T$ factorization
} \\

\vspace{4mm}

A.V. Kotikov$^a$,~~ A.V. Lipatov$^b$,~~ G. Parente$^c$,~~ N.P. Zotov$^d$

\vspace{4mm}
{}$^a$~BLThPh, JINR,
141980 Dubna, Russia,\\

{}$^b$~Dep. of Physics, MSU,
119899 Moscow, Russia,\\

{}$^c$~
Univ.
de Santiago de Compostela,
15706 Santiago de Compostela, Spain,\\

{}$^d$~INP, MSU,
119899 Moscow, Russia
\end{center}

\begin{abstract}
We present 
the perturbative parts of the structure functions
$F_2^c$ and $F_L^c$ for a gluon target having nonzero transverse
momentum squared at order $\alpha _s$.
The results of the double convolution (with respect to the Bjorken variable
$x$ and the
transverse momentum) of the perturbative part and the unintegrated
gluon densities are compared with HERA experimental data for
$F_2^c$.
The contribution from $F_L^c$ structure function ranges $10\div30\%$
of that of $F_2^c$ at the HERA kinematical range.
\end{abstract}


\pagestyle{plain}
\section{Introduction} \indent 

Recently there have been important new data on the charm structure function
(SF) $F_2^c$, of the proton from the H1 \cite{H1} and ZEUS \cite{ZEUS} 
Collaborations at HERA, which have probed the small-$x$ region down to 
$x=8\times 10^{-4}$ and $x=2\times 10^{-4}$, respectively. At these values 
of $x$, the charm contribution to the total proton SF, $F_2^p$, is found
to be around $25\%$, which is a considerably larger fraction than that found 
by the European Muon Collaboration at CERN \cite{EMC} at 
larger $x$, where it was only $\sim 1\%$ of $F_2^p$. Extensive 
theoretical analyses 
in recent years have generally served to confirm that 
the
$F_2^c$ data can be described through perturbative generation of charm 
within QCD (see, for example, the review in Ref. \cite{CoDeRo} and references 
therein).

We note, 
that perhaps more relevant analyses of the HERA data, where 
the $x$ values are quite small, are those based on BFKL dynamics
\cite{BFKL},
because the leading $ln(1/x)$ contributions are summed. The basic
dynamical quantity in BFKL approach is the unintegrated gluon distribution
$\vph_g(x,k^2_{\bot})$ 
\footnote{Hereafter
$p^{\mu}$ and $k^{\mu}$ are the hadron and the gluon 4-momentums, 
respectively,
and $q^{\mu}$ is the photon 4-momentum.}
($f_g$ is the (integrated) gluon distribution multiplied
by $x$ and $k_{\bot}$ is the transverse momentum)
 \begin{eqnarray}
f_{g}(x,Q^2) ~=~ f_{g}(x,Q^2_0) + \int^{Q^2}_{Q^2_0} dk^2_{\bot}
\; \vph_g(x,k^2_{\bot}), 
\label{1.1}
 \end{eqnarray}
which satisfies the BFKL equation.  
The integral is divergent at the lower limit and it leads 
to the necessity to use the difference $f_{g}(x,Q^2) - f_{g}(x,Q^2_0)$
with some nonzero $Q^2_0$.

Then, in the BFKL-like approach (hereafter the $k_t$-factorization
approach \cite{CaCiHa,CoEllis} is used)
the SFs $F^c_{2,L}(x,Q^2)$ are driven at small 
$x$ by gluons and are related in the following way to the unintegrated 
distribution $\vph_g(x,k^2_{\bot})$: 
\begin{eqnarray}
F^c_{2,L}(x,Q^2) ~=~\int^1_{x} \frac{dz}{z} \int dk^2_{\bot}
\; C^g_{2,L}(z,Q^2,m_c^2,k^2_{\bot})~ \vph_g(x/z, k^2_{\bot}), 
 \label{d1}
\end{eqnarray}

The functions $C^g_{2,L}(x,Q^2,m_c^2,k^2_{\bot})$ 
may be regarded as the structure 
functions of
the off-shell gluons with virtuality $k^2_{\bot}$ (hereafter we call them as
{\it hard structure functions}\footnote{This 
notation reflects the fact that structure functions  
$F_{2,L}^c$ connect with the functions $C_{2,L}^g$ at the same form 
as cross-sections connect with hard ones (see \cite{CaCiHa,CoEllis}).}). 
They are described by the quark box (and
crossed box) diagram contribution to the photon-gluon interaction 
(see Fig. 1 in \cite{KoLiPaZo}). 

The purpose of the paper is to present the results for
these hard SF
$C^g_{2,L}(x,Q^2,m_c^2,k^2_{\bot})$ and to analyze experimental data for 
$F^c_{2}(x,Q^2)$ by applying Eq. (\ref{d1}) with different sets of
unintegrated gluon densities (see Ref. \cite{LiZo,Andersson})
and to give predictions for the longitudinal SF $F^c_{L}(x_B,Q^2)$.

It is instructive to note that
the results should be  
similar to those  of the
photon-photon scattering process.
The corresponding QED contributions have been calculated many years ago
in Ref. \cite{BFaKh} (see also the beautiful review in Ref. \cite{BGMS}). 
Our results 
have been calculated independently in \cite{KoLiPaZo} 
(based on approaches of \cite{KaKo,KoTMF})
and they are in full agreement with 
\cite{BFaKh}. 
However, we hope
that our formulas which are given in a 
simpler form 
could be useful for others.

\section{Hard structure functions} \indent

The gluon polarization tensor (hereafter
the indices $\alpha$ and $\beta$ are connected with gluons),
which gives the main contribution at high energy limit, has the form:
\bea
\hat P^{\alpha\beta}_{BFKL}~=~
\frac{k_{\bot}^{\alpha}k_{\bot}^{\beta}}{k_{\bot}^2}
~=~\frac{x^2}{-k^2}
p^{\alpha}p^{\beta}~=~ -\frac{1}{2}
\frac{1}{\tilde \beta^4} \left[\tilde \beta^2 g^{\alpha \beta} 
-12 bx^2 \frac{q^{\alpha}q^{\beta}}{Q^2} \right],
\label{1dd}
\eea
where $\tilde \beta^2=1-4bx^2,~~b=-k^2/Q^2 \equiv  k_{\bot}^2/Q^2 >0.$

Contracting the corresponding photon projectors, we have: 
 \begin{eqnarray}
\z C^g_{2}(x) = 
\frac{{\cal K}}{\tilde \beta^2 }
\;
\left[
f^{(1)}_{BFKL} + 
\frac{3}{2\tilde \beta^2}\; f^{(2)}_{BFKL} \right],~
C^g_{L} (x) = 
\frac{{\cal K}}{\tilde \beta^2 }
\;
\left[
4bx^2 f^{(1)}_{BFKL} + 
\frac{(1+2bx^2)}{\tilde \beta^2}\; f^{(2)}_{BFKL} \right]
\label{3}\\
\z f^{(1)}_{BFKL} = 
\frac{1}{\tilde\beta^4}
\left[ \tilde \beta^2 f^{(1)}  ~-~
3bx^2 \tilde f^{(1)}\right],~~~
f^{(2)}_{BFKL} ~=~
\frac{1}{\tilde\beta^4}
\left[ \tilde \beta^2 f^{(2)}  ~-~
3bx^2 \tilde f^{(2)} \right],
\nonumber
 \end{eqnarray}
where the normalization factor 
${\cal K} =  e_c^2 \; \alpha_s(Q^2)/(4\pi)\; x$, 
~$a=m^2_c/Q^2$~, $e_c=2/3$ is charm quark charge 
and
\begin{eqnarray}
f^{(1)} &=& -2 \beta \Biggl[ 1 - \biggl(1-2x(1+b-2a) \; [1-x(1+b+2a)] 
\biggr) \; f_1  
\nonumber \\
    &+& (2a-b)(1-2a)x^2 \; f_2  \Biggr],
 \label{5}\\
f^{(2)} &=& 8x\; \beta \Biggl[(1-(1+b)x)  
-2x \biggl(bx(1-(1+b)x)(1+b-2a) + a\tilde \beta^2 \biggr)\; f_1  
\nonumber \\
    &+& bx^2(1-(1+b)x) (2a-b) \; f_2  \Biggr],
\label{6} \\
\tilde f^{(1)} &=& - \beta \Biggl[ \frac{1-x(1+b)}{x}  
-2 \biggl(x(1-x(1+b))(1+b-2a) +a \tilde \beta^2 \biggr) \; f_1  
\nonumber \\
    &-& x(1-x(1+b))(1-2a) \; f_2  \Biggr],
 \label{5dd}\\
\tilde f^{(2)} &=& 4 \; \beta ~(1-(1+b)x)^2 \Biggl[2   
- (1+2bx^2)\; f_1  
-bx^2 \; f_2  \Biggr],
\label{6dd} 
\end{eqnarray}
with
$$
\beta^2=1-\frac{4ax}{(1-(1+b)x)},~~~~~f_1=\frac{1}{\tilde \beta \beta} \;
\ln\frac{1+\beta \tilde \beta}{1-\beta \tilde \beta},
~~~~~f_2=\frac{-4}{1-\beta^2 \tilde \beta^2}$$

For the important regimes when
$k^2= 0$, $m^2_c = 0$ and $Q^2 = 0$, the 
results are coincided with ones of Refs. \cite{CaCiHa,Witten}.
Notice that our results in Eq. (\ref{3})  should
be also
agree with those in Ref. \cite{BMSS} but the direct comparison is 
quite difficult because 
the structure of their results is quite cumbersome (see Appendix A in 
Ref. \cite{BMSS}). 
We have found numerical agreement in the case of 
$F_2(x,Q^2)$ for several types of unintegrated gluon distributions
(see Fig.4 in \cite{KoLiPaZo}).

\section{Comparison with $F_2^c$ experimental data and 
predictions for $F_L^c$ } \indent

With the help of the results obtained in the previous Section
we have analyzed HERA data for SF $F_2^c$ from ZEUS \cite{ZEUS} 
collaboration.

Notice that in Ref. \cite{CaCiHa} 
the $k_{\bot}^2$-integral in the r.h.s. of Eq.(\ref{d1}) has been evaluated
using the BFKL results for the Mellin transform of the
unintegrated gluon distribution and the Wilson coefficient functions have 
been calculated for the full perturbative series at asymptotically small
$x$ values. Since we would like to analyze experimental data for the SF
$F_2^c$, we have an interest to obtain results at quite broad range of 
small $x$ values. For the reason we need in a parameterization
of unintegrated gluon distribution function.

We consider two different parametrizations for the
unintegrated gluon distribution (see ~\cite{LiZo}):
the Ryskin-Shabelski (RS) one ~\cite{RS}
and the Blumlein one 
\cite{BL}. 
\begin{figure}
%
\begin{center}
\vskip -1cm
\epsfig{figure=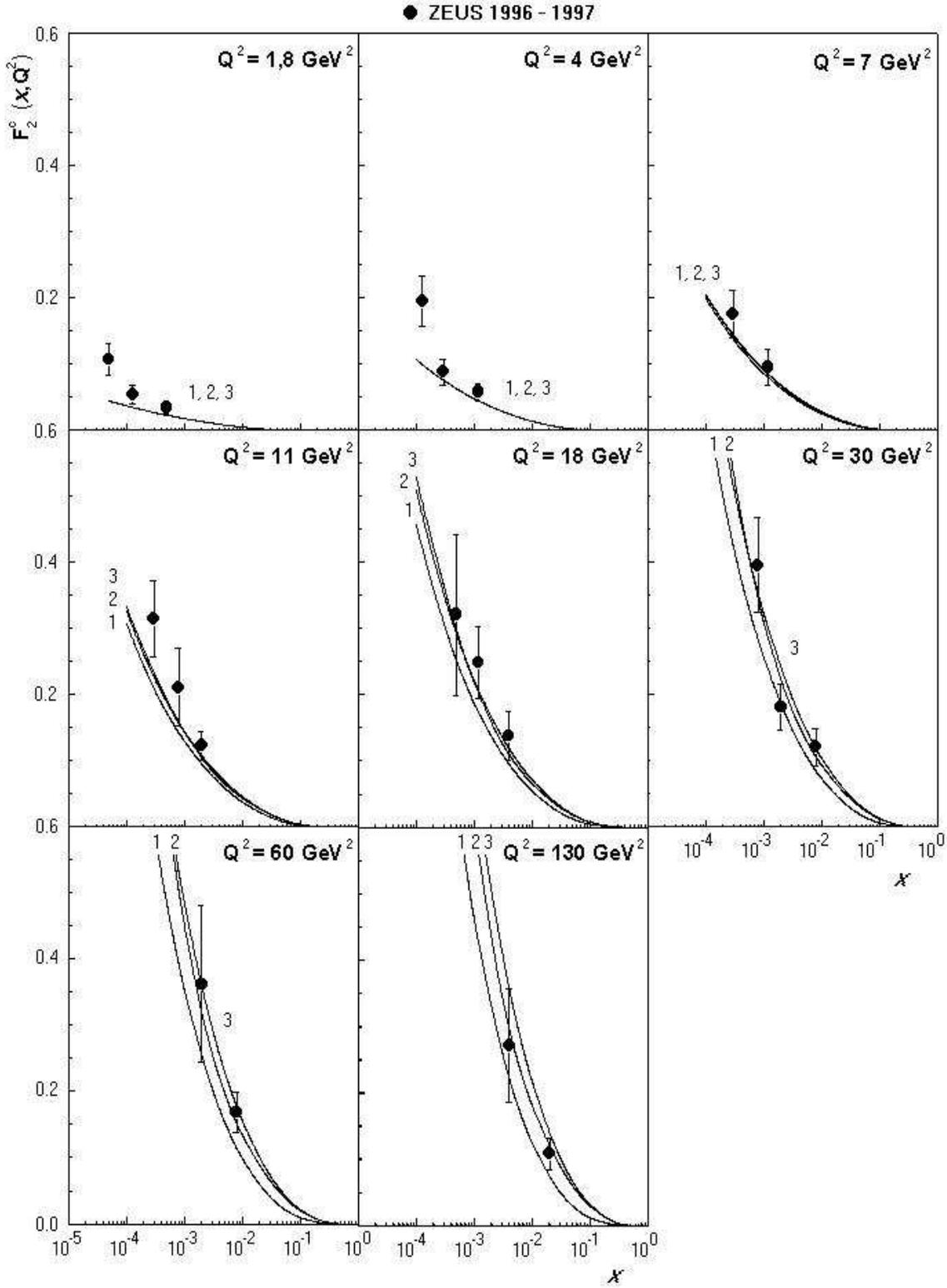,width=16.5cm,height=21.8cm}
\end{center}
\vskip -1cm
\caption{The structure function $F_2^c(x,Q^2)$ as a function of $x$ for
different values of $Q^2$ compared to ZEUS data~\cite{ZEUS}.
Curves 1, 2 and 3 correspond to SF obtained
in the standard parton model with the GRV ~\cite{GRV}
gluon density at the leading order approximation and to SF
obtained in the $k_T$ factorization approach with RS~\cite{RS} 
and Blumlein (at $Q_0^2 = 4$ GeV$^2$)~\cite{BL} parameterizations of 
unintegrated gluon distribution.}
\end{figure}

In Fig. 1 
we show  the SF $F_2^c$ as a function $x$ for different values
of $Q^2$  in comparison with ZEUS \cite{ZEUS} 
experimental data.
 We see that at large $Q^2$ ($Q^2 \geq 10$ GeV$^2$)
the SF $F_2^c$
obtained in the $k_T$ factorization approach is higher than the SF obtained
in the standard parton model with the GRV \cite{GRV}
gluon density at the LO approximation 
(see curve 1) 
and has a more rapid growth in comparison with the standard parton model
results, especially at $Q^2 \sim 130$ GeV$^2$~\cite{LZ}.
At $Q^2 \leq 10$ GeV$^2$ the predictions from
perturbative QCD (in GRV approach)
and those based on the $k_T$ factorization approach are very similar
\footnote{This fact is due to the quite large value of $Q^2_0=4$ 
GeV$^2$ chosen here.} and show a disagreement with data 
below $Q^2 = 7$ GeV$^2$
\footnote{A similar disagreement with data at $Q^2 \leq 2$ GeV$^2$
has been observed for the complete structure function $F_2$
(see, for example, the discussion in Ref. \cite{Q2evo}
and reference therein). 
We note that the insertion of higher-twist corrections in the framework
of usual perturbative QCD improves the agreement with data
(see Ref. \cite{HT}) at quite low values of $Q^2$.}.
Unfortunately the available experimental data do not permit yet
to distinguish 
the $k_T$ factorization effects from
those due to boundary conditions~\cite{RS}.

The 
results for the SF  $F_L^c$ obtained in perturbative QCD and 
from the $k_T$ factorization approach are quite similar to the $F_2^c$ case
discussed above.
The ratio $ R^c = F_L^c/F_2^c$ is 
shown in Fig. 2. We see that $ R^c \approx 0.1 \div 0.3 $ in a wide 
region of $Q^2$.
The estimation of $R^c$ is very close to the results for $R=F_L/(F_2-F_L)$
ratio (see Refs. \cite{KoPaFL}-\cite{CCFRr}).
We would like to note that these values of $ R^c $
contradict the estimation
obtained in Refs.~\cite{ZEUS, H1}. The effect of $ R^c $
on the corresponding differential cross-section
should be considered in the extraction of $F_2^c$
from future more precise measurements.

\begin{figure}
\vskip -1cm
\epsfig{figure=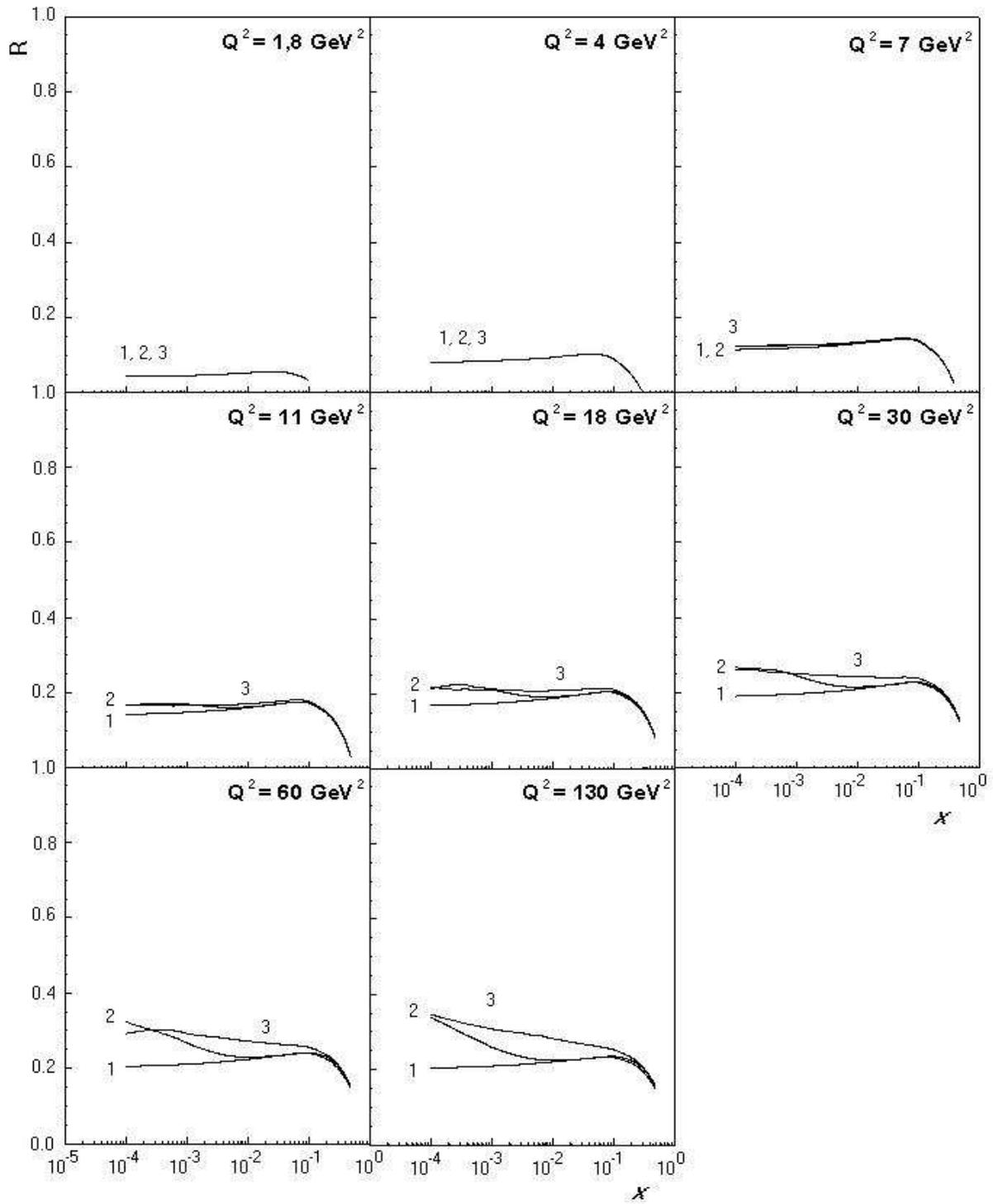,width=17.5cm,height=20.8cm}
\vskip -0.5cm
\caption{The ratio  $R^c  = F_L^c(x,Q^2)/F_2^c$ as a function of $x$ 
for different values of $Q^2$.
Curves 1, 2 and 3 are as in Fig. 1.}
\label{fig8}
\end{figure}

For the ratio $ R^c $ we found quite flat $x$-behavior at low $x$ 
in the low $Q^2$ region (see Fig. 2), where approaches based on perturbative
QCD and on $k_T$ factorization give similar predictions 
(see Fig 1).
It is in agreement with the corresponding behaviour of the ratio
$R=F_L/(F_2-F_L)$ (see Ref. \cite{KoPaFL}) at quite large values of $\Delta_P$
\footnote{The behaviour is in agreement with previous studies 
\cite{TMF,KoPaFL}. Note that
at small values of $\Delta_P$, i.e. when $x^{-\Delta_P} \sim Const$,
the ratio $R$ has got strong negative NLO corrections 
(see Refs. \cite{Keller,resum})
and tends to zero at $x \to 0$ after some resummation done in \cite{resum}.}
($\Delta_P > 0.2-0.3$).
The low $x$ rise of $ R^c $ at high $Q^2$ disagrees with early
calculations \cite{KoPaFL} in the framework of perturbative QCD.
It could be due to the small $x$ resummation, which is important at high $Q^2$
(see Fig 1).
We plan to study this effect in future.

\vspace{-0.3cm}
\section{Conclusions} \indent

We have presented
the results for
the perturbative
parts of the SFs
$F_2^c$ and $F_L^c$ for a gluon target 
having nonzero momentum squared, in the process 
of photon-gluon fusion.

We have applied the
results in the framework of $k_T$ factorization approach
to the analysis of
present data for the charm contribution to $F_2$ 
and we have given the predictions for $F_L^c$.
The analysis has been performed with
two parameterizations of unintegrated gluon
distributions  for comparison. We have found good agreement
of our results
with experimental HERA data for $F_2^c$,
except at low $Q^2$ ($Q^2 \leq 7$ GeV$^2$)
\footnote{It must be noted that the cross section of inelastic $c\bar c-$
and $b\bar b-$pair photoproduction at HERA are described by the
Blumlein parametrization at a smaller value of $Q_0^2$ ($Q_0^2 = 1$ GeV$^2$)
(see \cite{LiZo}).}. 
We have also obtained
quite large contribution of the SF $F_L^c$ at low $x$  and
high $Q^2$ ($Q^2 \geq 30$ GeV$^2$).

We would like to note the good agreement between our results for $F_2^c$
and the ones obtained in Ref. \cite{Jung} by Monte-Carlo studies. Moreover, we
have also good agreement with fits of H1 and ZEUS data for $F_2^c$
(see recent reviews in Ref. \cite{Wolf} and references therein)
based on perturbative
QCD calculations. But unlike to these fits,
our analysis uses universal unintegrated gluon distribution, which gives
in the simplest way the main contribution to the cross-section in the
high-energy limit.

It could be also very useful to evaluate the complete $F_2$ itself and 
the derivatives of $F_2$ with respect to the 
logarithms of $1/x$ and $Q^2$ with our expressions using the unintegrated
gluons.

The consideration of the SF $F_2$ in the framework of the leading-twist
approximation of perturbative QCD (i.e. for ``pure'' perturbative QCD)
leads to very good agreement (see Ref. \cite{Q2evo} and references therein) 
with HERA
data at low $x$ and $Q^2 \geq 1.5$ GeV$^2$. The agreement improves
at lower $Q^2$ when higher twist terms are 
taken into account \cite{HT}. As
it has been studied in Refs. \cite{Q2evo,HT}, the SF $F_2$  at low 
$Q^2$ is sensitive to the
small-$x$ behavior of quark distributions.
Thus, our future 
analysis of $F_2$ in a broader $Q^2$ range in the framework
of $k_T$ factorization 
should require the incorporation of parametrizations for unintegrated
quark densities, introduced recently (see Ref. \cite{Ryskin} and references
therein).

The 
SF $F_L$ is
very interested too (see \cite{KoLiZo}).
The structure function $F_L$ depends strongly on the gluon distribution
(see, for example, Ref. \cite{CSI...,method}), which in turn is determined 
\cite{Prytz} by the derivative $dF_2/d\ln Q^2$. Thus, in the framework of 
perturbative QCD at low $x$ the relation between  $F_L$, $F_2$ and
$dF_2/d\ln Q^2$ could be violated by non-perturbative contributions,
which were expected to be important in the $F_L$ case (see Refs. 
\cite{BaCoPe,PaKoKri}).
The application in \cite{KoLiZo} of present analysis to $F_L$ gives a 
``non pure'' perturbative QCD predictions for the structure function 
which are in good agreement with  
data \cite{CCFRr} and with the ``pure'' perturbative results of 
Ref. \cite{KoPaFL}, that determines a small value of nonperturbative
corrections to SF $F_L$ at low $x$ and quite large $Q^2$ values.

%
%

\vskip 0.2cm
{\bf Acknowledgments.}~~Authors
 would like to express their sincerely thanks to the Organizing
  Committee of the XVIth International Workshop ``High Energy Physics
and Quantum Field Theory'' for the kind invitation,  
the financial support
 at  such remarkable Conferences, and 
 for fruitful discussions.

A.V.K. was supported in part
by Alexander von Humboldt fellowship and INTAS  grant N366.
G.P. acknowledges the support of Galician research funds
(PGIDT00 PX20615PR) and Spanish CICYT (FPA2002-01161).
N.P.Z. also acknowledge the support of Royal Swedish Academy of
Sciences.






\vspace{-0.5cm}
\begin{thebibliography}{99}


\bibitem{H1} H1 Coll.: S. Aid et al.,
Z.Phys. C72 (1996) 593; Nucl.Phys. B545 (1999) 21.
\vspace{-0.2cm} 
\bibitem{ZEUS} ZEUS Coll.: J. Breitweg et al.,
Phys.Lett. B407 (1997)402;
%
 Eur.Phys.J. C12 (2000)35.
%
\vspace{-0.2cm} 
%
\bibitem{EMC} EM Coll.: J.J. Aubert et al.,
Nucl.Phys. B213 (1983) 31;
Phys.Lett. B94 (1980) 96;
B110 (1983) 72.
\vspace{-0.2cm} 
%
\bibitem{CoDeRo} A.M. Cooper-Sarkar et al.,
Int.J.Mod.Phys. A13 (1998) 3385.
\vspace{-0.2cm} 
%
\bibitem{BFKL} L.N. Lipatov, Sov.J.Nucl.Phys.
23 (1976) 338;
E.A. Kuraev et al.,
Sov.Phys.JETP 44 (1976) 443, 45 (1977) 199;
Ya.Ya. Balitzki, L.N. Lipatov, Sov.J.Nucl.Phys.
28 (1978) 822;
 L.N. Lipatov, JETP 63 (1986) 904.
\vspace{-0.2cm} 
%
\bibitem{CaCiHa} S. Catani et al., 
Phys.Lett. B242 (1990) 97;
Nucl.Phys. B366 (1991) 135.
\vspace{-0.2cm} 
%
\bibitem{CoEllis} J.C. Collins, R.K. Ellis,
Nucl.Phys. B360 (1991) 3; 
E.M. Levin et al., 
Sov. J. Nucl. Phys. 53 (1991) 657.
\vspace{-0.2cm} 
%
%
%
\bibitem{KoLiPaZo}
A.V. Kotikov et al.,
Preprint US-FT/7-01
(hep-ph/0107135).
\vspace{-0.2cm} 
%
\bibitem{LiZo}  A. V. Lipatov et al., 
Mod.Phys.Lett. A15 (2000) 695;
1727.
\vspace{-0.2cm} 
%
\bibitem{Andersson}  Bo Andersson et al., hep-ph/0204115.
\vspace{-0.2cm} 
%
\bibitem{BFaKh} V.N. Baier et al.,
JETP 50 (1966) 156;
V.G. Zima, Yad.Fiz. 16 (1972) 1051.
\vspace{-0.2cm} 
%
\bibitem{BGMS} V.M. Budnev et al.,
Phys.Rept. 15 (1975) 181.
\vspace{-0.2cm} 
%
\bibitem{KaKo} D.I. Kazakov,  A.V. Kotikov,
Theor.Math.Phys. 73 (1987) 1264;
Nucl.Phys. B307 (1988) 721; 
Nucl.Phys. B345 (1990) 299.
\vspace{-0.2cm} 
%
%
\bibitem{KoTMF}
 A.V. Kotikov, 
Theor.Math.Phys. 78 (1989) 134;
Phys.Lett. B375 (1996) 240;
%
Phys.Lett. 
B254 (1991) 158; 
Phys.Lett.
B259 (1991) 314;
Phys.Lett. B267 (1991) 123;
in: {\it Proc. of the XXXV
Winter School}, Repino, S'Peterburg, 2001 (hep-ph/0112347).
\vspace{-0.2cm} 
%
\bibitem{Witten} E. Witten,
Nucl.Phys. B104 (1976) 445;
M. Gluck, E. Reya,
Phys.Lett. B83 (1979) 98;
%
F.M. Steffens et al., 
Eur.Phys.J. C11 (1999) 673.
\vspace{-0.2cm} 
%
\bibitem{BMSS}  G. Bottazzi et al., 
JHEP 9812 (1998) 011.
\vspace{-0.2cm} 
%
%
\bibitem{RS} M.G.~Ryskin, Yu.M.~Shabelski,  Z.Phys. C61
 (1994) 517; C66 (1995) 151. 
\vspace{-0.2cm} 
%
\bibitem{BL}  J. Blumlein, 
Preprints DESY 95-121 (hep-ph/9506403); DESY 95-125 (hep-ph/9506446).
\vspace{-0.2cm} 
%
\bibitem{GRV} M. Gluck et al.,
Z. Phys. C67 (1995) 433.
\vspace{-0.2cm} 
%
\bibitem{LZ} A.V. Lipatov, N.P. Zotov, {\it in} Proc. of the 8th Int. 
Workshop 
DIS 2000 (2000), World Scientific,  p. 157.
\vspace{-0.2cm} 
\bibitem{Q2evo} A.V. Kotikov, G. Parente,
Nucl.Phys. B549 (1999) 242; 
hep-ph/0010352; 
{\it in} Proc. of the Int. Conference PQFT98 (1998), Dubna
(hep-ph/9810223);
{\it in} Proc. of the 8th Int. Workshop on Deep Inelastic
Scattering, DIS 2000 (2000), Liverpool, p. 198 
(hep-ph/0006197);
Preprint US-FT/3-02 (hep-ph/0207276).
\vspace{-0.2cm} 
%
%
\bibitem{HT} A.V. Kotikov, G. Parente, 
 {\it in} Proc. Int. Seminar Relativistic Nuclear Physics and Quantum 
Chromodynamics (2000), Dubna
(hep-ph/0012299); 
{\it in} Proc. of the 9th Int. Workshop on Deep Inelastic
Scattering, DIS 2001 (2001), Bologna (hep-ph/0106175).
\vspace{-0.2cm} 
%
%
\bibitem{KoPaFL} A.V. Kotikov, JETP 80 (1995) 979;
A.V. Kotikov, G. Parente, 
{\it in} Proc. Int. Workshop on Deep Inelastic
Scattering 
(1996), Rome, p. 237 
(hep-ph/9608409);
Mod.Phys.Lett. A12 (1997) 963;
JETP 85 (1997) 17;
hep-ph/9609439.
\vspace{-0.2cm} 
%
%
\bibitem{Thorne} R.S. Thorne, 
Phys.Lett. B418 (1998) 371.
\vspace{-0.2cm} 
%
\bibitem{CCFRr}  H1 Coll.: S. Aid et al., Phys.Lett. B393 (1997) 452;
CCFR/NuTeV Coll.: U.K. Yang et al., hep-ex/0010001;
A. Bodek, hep-ex/00105067
\vspace{-0.2cm} 
%
\bibitem{TMF} A.V. Kotikov et al., 
Theor. Math. Phys. {\bf 84} (1990) 744;
Theor. Math. Phys. {\bf 111} (1997) 442;
A.V.~Kotikov, Phys. Atom. Nucl. 56 (1993) 1276;
%
L.~L.~Jenkovszky et al., 
Sov.\ J.\ Nucl.\ Phys.\  {\bf 55} (1992) 1224;.
JETP Lett.\  {\bf 58} (1993) 163;
Phys.\ Lett.\ B {\bf 314}(1993) 421.
\vspace{-0.2cm} 
%
\bibitem{Keller} S. Keller et al., 
Phys.Lett. B270 (1990) 61;
L.H. Orr, W.J. Stirling, Phys.Rev.Lett. B66 (1991) 1673;
E. Berger, R. Meng, Phys.Lett. B304 (1993) 318.
\vspace{-0.2cm} 
%
\bibitem{resum}
A.V. Kotikov, JETP Lett. 59 (1994) 1; Phys.Lett. B338 (1994) 349. 
\vspace{-0.2cm} 
%
\bibitem{Jung} H. Jung, Nucl.Phys.(Proc.Suppl.) 79 (1999) 429.
\vspace{-0.2cm} 
%
\bibitem{Wolf} G. Wolf, Preprint DESY 01-058 (hep-ex/0105055).
\vspace{-0.2cm} 
%
\bibitem{Ryskin} M.A. Kimber et al., 
Phys.Rev. D63 (2001) 114027.
\vspace{-0.2cm} 
%
%
\bibitem{KoLiZo}
S. Catani and F. Hautmann,
Nucl.Phys. {\bf B427} (1994) 475;
A.V. Kotikov et al., 
hep-ph/0207226.
\vspace{-0.2cm} 
%
\bibitem{CSI...} A.M. Cooper-Sarkar et al., 
Z.Phys. C39 (1988) 281;
A.V. Kotikov, G. Parente, in: {\it Proc. of 
Madrid Workshop on Low x Physics} (1997) pp. 71-76  
(hep-ph/9710252). 
\vspace{-0.2cm} 
%
\bibitem{method} A.V. Kotikov, Phys.Atom.Nucl. 57 (1994) 133;
Phys.Rev. D49 (1994) 5746.
\vspace{-0.2cm} 
%
\bibitem{Prytz} K. Prytz,  Phys.Lett. B311 (1993) 286;
A.V. Kotikov, JETP Lett. 59 (1994) 667;
A.V. Kotikov, G. Parente,  Phys.Lett. B379 (1996) 195. 
\vspace{-0.2cm} 
%
\bibitem{BaCoPe} J. Bartels et al.,
Eur.Phys.J. C17 (2000) 121;
\vspace{-0.2cm} 
%
\bibitem{PaKoKri}
A.V.~ Kotikov at al., 
Z. Phys. C58 (1993) 465;
G.~Parente et al., 
Phys.Lett. B333 (1994) 190; 
A.V. Kotikov, V.G. Krivokhijine, Dubna preprint E2-2001-190 
(hep-ph/0108224);
in: {\it Proc. of the XVIth International Workshop 
``High Energy Physics and Quantum Field Theory``} (2001), Moscow;
(hep-ph/0206221);
in: {\it Proc. of the Int.
Workshop ``Renormalization Group 2002''} (2002), High Tatras, Slovakia
(hep-ph/0207222).
in: {\it Proc. of the X Int. Workshop on Deep Inelastic Scattering
  (DIS2002)} (2002), Cracow
(hep-ph/0208188).
\vspace{-0.2cm} %
\end{thebibliography}
\end{document}